\documentclass[12pt]{article}

\begin{document}

\begin{center}
{\Large {\bf Anomalous dimensions of Wilson operators \\[5mm]
in N=4 SYM 
Theory.}} \\[0pt]

\vspace{1.5cm} {\large \ A.~V.~Kotikov } \\[0pt]
\vspace{0.5cm} {\em Bogoliubov Laboratory of Theoretical Physics \\[0pt]
Joint Institute for Nuclear Research\\[0pt]
141980 Dubna, Russia }\\[0pt]

\vspace{0.5cm} {\large \ L.~N.~Lipatov and V.~N.~Velizhanin 
\footnote[2]{velizh@thd.pnpi.spb.ru}}\\[0pt]
\vspace{0.5cm} {\em Theoretical Physics Department\\[0pt]
Petersburg Nuclear Physics Institute\\[0pt]
Orlova Roscha, Gatchina\\[0pt]
188300, St. Petersburg, Russia }\\[0pt]
\end{center}

\vspace{1cm}

\begin{center}
{\bf Abstract}
\end{center}

We present the results of two-loop calculations of the anomalous
dimension matrix for the Wilson twist-2 operators in the N=4
Supersymmetric Yang-Mills theory for polarized and unpolarized
cases. This matrix can be transformed to a triangle form by the
same similarity transformation as in the leading order. The
eigenvalues of the anomalous dimension matrix are expressed in
terms of an universal function with its argument shifted by
integer numbers. In the conclusion we discuss relations
between the weak and strong coupling regimes in the framework of the AdS/CFT
correspondence.


\newpage

Parton distributions in QCD satisfy the Balitsky-Fadin-Kuraev-Lipatov
(BFKL) \cite{BFKL} and Dokshitzer-Gribov-Lipatov-Altarelli-Parisi (DGLAP)
\cite{DGLAP,Dokshitzer} equations. Next-to-leading corrections to the
BFKL equation were calculated only recently~\cite{next}. It is natural
to generalize these equations to the supersymmetric case (see Refs.~\cite{Dubna,
KL, KL01} and references therein). Indeed, the supersymmetric field
theories have a number of amazing properties, such as a cancellation of
quadratic divergencies and non-renormalization theorems for interaction
terms in the lagrangian.  Moreover, the supersymmetry is an excellent
technical playground for QCD.  For example, the empirically established
Dokshitzer relation~\cite {Dokshitzer} among elements of the leading
order anomalous dimension matrix in the N=1 supersymmetric limit
provides a non-trivial check of results of higher order calculations.
Another interesting example is the relation between the BFKL and DGLAP
equations in the N=4 Supersymmetric Yang-Mills (SYM)
theory~\cite{Dubna,KL}. In this model one can obtain the anomalous
dimensions of the multiplicatively renormalizable twist-2 operators from the
eigenvalues of the BFKL kernel~\cite{KL01}. These operators are certain
linear combinations of the Wilson operators appearing in the theoretical
description of the deep-inelastic $ep$ scattering~\cite{Dubna,KL} (note,
that in the N=4 SYM the beta function is zero and the Bjorken scaling for
structure functions is strongly violated). Using some assumptions the
authors of Ref.~\cite{KL} derived also an expression for the universal
anomalous dimension for the N=4 model in the two-loop approximation.
Moreover, the eigenvalues of the anomalous dimension matrices in the
polarized and unpolarized cases were obtained from this universal anomalous
dimension by an appropriate integer shift of its argument. In this paper we
present the results of direct two-loop calculations of these matrices in the
N=4 SYM theory.

Now the anomalous dimensions of twist-2 operators in QCD are known up to two
loops both for the unpolarized~\cite{AnomDim:UnPol,HVN} and polarized~\cite
{MVN,Vogelsang:1996im} cases. In the N=4 SYM theory~\cite{Brink:1976bc}
there are one gluon $g$, four Majorana fermions $q$, three scalars and three
pseudoscalars which can be unified in three complex scalars $\varphi $. All
particles belong to the adjoint representation of the gauge group
$SU(N_{c})$. The transition from QCD to the N=4 SYM theory can be performed if one puts
in the final expressions $C_{A}=C_{F}=N_{c}$, $T_{f}=2N_{c}$ (the last
substitution follows from the fact, that each gluino $q_{i}$ from four
Majorana particles gives a half of the Dirac spinor contribution).
Furthermore, one should take into account the diagrams with virtual scalars
in the polarized structure functions and the graphs with external scalars in
the non-polarized distributions. In the last case the anomalous dimension
matrix extends to $3\times 3$. Below we calculate the anomalous dimensions
of the following gauge-invariant twist-2 operators:
\begin{eqnarray}
{\cal O}^g_{\mu_1,...,\mu_j}&=&\hat S
G_{\rho\mu_1}D_{\mu_2}D_{\mu_3}...D_{\mu_{j-1}}G_{\rho\mu_j}\,, \\
{\tilde {{\cal O}}}^g_{\mu_1,...,\mu_j}&=&\hat S
G_{\rho\mu_1}D_{\mu_2}D_{\mu_3}...D_{\mu_{j-1}}{\tilde G}_{\rho\mu_j}\,, \\
{\cal O}^q_{\mu_1,...,\mu_j}&=&\hat S\bar\Psi \gamma_{\mu_1}
D_{\mu_2}...D_{\mu_j}\Psi \,, \\
{\tilde {{\cal O}}}^q_{\mu_1,...,\mu_j}&=&\hat S\bar\Psi \gamma_5
\gamma_{\mu_1} D_{\mu_2}...D_{\mu_j}\Psi \,, \\
{\cal O}^{\varphi}_{\mu_1,...,\mu_j}&=&\hat S\bar\Phi D_{\mu_1}
D_{\mu_2}...D_{\mu_j}\Phi \,,
\end{eqnarray}
where $D_{\mu}$ are covariant derivatives; the spinor $\Psi$ and field
tensor $G_{\rho\mu}$ describe gluinos and gluons, respectively, and $\Phi$
is the complex scalar field appearing in the N=4 supersymmetric model. The
symbol $\hat S$ implies a symmetrization of the tensor in the Lorenz indices
$\mu_1,...,\mu_j$ and a subtraction of its traces. The anomalous dimension
matrices can be written as follows for the unpolarized 
\begin{equation}
\gamma_{{\bf unpol}}=
\begin{array}{|ccc|}
\gamma_{gg} & \gamma_{gq} & \gamma_{g\varphi} \\
\gamma_{qg} & \gamma_{qq} & \gamma_{q\varphi} \\
\gamma_{\varphi g} & \gamma_{\varphi q} & \gamma_{\varphi\varphi}
\end{array}
\end{equation}
and polarized cases
\begin{equation}
\gamma_{{\bf pol}}=
\begin{array}{|cc|}
{\tilde \gamma}_{gg} & {\tilde \gamma}_{gq} \\
{\tilde \gamma}_{qg} & {\tilde \gamma}_{qq}
\end{array}
\,.
\end{equation}
Note, that in the super-multiplet of twist-2 operators there are also
operators with fermion quantum numbers and operators anti-symmetric in two
Lorentz indices~\cite{BFKL2}.

Our approach is similar to that of Refs.~\cite{HVN,MVN}. In particular we
calculated unrenormalized matrix elements of the partonic operators
sandwiched between the scalars, fermion and gluon states 
\footnote[4]{ 
For the calculations we used the program DIANA~\cite
{DIANA}, which calls QGRAF~ \cite{QGRAF} for the generation of Feynman
diagrams, and the package MINCER~ \cite{MINCER} for FORM~\cite{FORM} for the
evaluation of two-loop diagrams.}
and the anomalous dimensions were extracted from the expansion
of the matrix elements through the renormalization group coefficients, with
the condition, that the renormalized matrix elements satisfy the
Callan-Symanzik equations. In our calculations we used the modified minimal
subtraction scheme (${\overline{{\rm MS}}}$). Because this scheme violates
the supersymmetry, the results were transformed to the dimensional reduction
scheme (${\overline{{\rm DR}}}$) \cite{Siegel}, explicitly preserving
supersymmetry at least in the two-loop level. For this purpose we used the
same procedure as in Ref.~\cite{Antoniadis:1981zv}. Namely, the difference
of two-loop results in ${\overline{{\rm MS}}}$~and 
${\overline{{\rm DR}}}$-schemes was related to the difference of the finite 
contributions of the corresponding one-loop results.

In the polarized case one needs an appropriate choice for the 
$\gamma _{5}$-prescription. 
Our procedure is analogous to that of Ref.~\cite{MVN}, which
based on ''reading point'' method~\cite{KKS}. To begin with, in each trace
of the $\gamma $-matrix product, we pushed $\gamma_{5}$ to the right hand
side using the property of the trace cyclicity. After that we simplified in
a straightforward way the product of $\gamma$-matrices leaving the $\gamma
_{5}$-matrix untouched and used the relation ${\rm Tr}\,\gamma _{\mu }\gamma
_{\nu }\gamma _{\rho }\gamma _{\sigma }\gamma _{5}=-4i\epsilon _{\mu \nu
\rho \sigma }$. Then the integration over the loop momenta in the space-time
dimension $D=4-2\varepsilon$ was performed and the contraction between two
Levi-Civita tensors (second one appearing from the projector) in four
dimensions was done. In the end we introduced an additional renormalization
constant to restore the anticommutativity of $\gamma _{5}$ with other 
$\gamma $-matrices in an accordance with Ref.~\cite{Larin}.

The final two-loop result for the elements of the anomalous dimension matrix
in N=4 SYM theory in the ${\overline{{\rm DR}}}$-scheme has the following
form (multiplied by $\alpha _{s}^{2}N_{c}^{2}/(4\pi )^{2}$) in the
unpolarized case (for even $j$)
\begin{eqnarray}
\gamma _{gg}^{(1)}(j)
&=&\frac{-500}{9(j-1)}-\frac{16}{j^{3}}+\frac{72}{j^{2}}
+\frac{140}{3\,j}+\frac{24}{(j+1)^{2}} 
-\frac{236}{3(j+1)}-\frac{16}{(j+2)^{3}}+\frac{176}{3\,(j+2)^{2}}\nonumber \\
&+& \frac{788}{9(j+2)}  
-16\,K(j-1)+16\,K(j)-16\,K(j+1)  
+16\,K(j+2)+{\hat{Q}}(j)\,,  \nonumber \\
\gamma _{gq}^{(1)}(j)
&=&\frac{-500}{9(j-1)}-\frac{16}{j^{3}}+\frac{72}{j^{2}}
+\frac{140}{3\,j}-\frac{22}{3(j+1)}  
+\frac{32}{3\,(j+2)^{2}}+\frac{152}{9(j+2)} \nonumber \\
&-&16\,K(j-1)+16\,K(j)-8\,K(j+1)\,,  \nonumber \\
\gamma _{g\varphi }^{(1)}(j)
&=&\frac{-500}{9(j-1)}-\frac{16}{j^{3}}+\frac{72}{j^{2}}
+\frac{140}{3\,j}-\frac{8}{(j+1)^{2}} 
+\frac{16}{j+1}-\frac{16}{3\,(j+2)^{2}}
-\frac{64}{9(j+2)}  \nonumber \\
&-&16\,K(j-1)+16\,K(j) \,,  \nonumber\\
\gamma _{qg}^{(1)}(j)
&=&\frac{320}{9(j-1)}+\frac{32}{j^{3}}-\frac{96}{j^{2}}%
+\frac{8}{3\,j}-\frac{96}{(j+1)^{2}}  
+\frac{944}{3(j+1)}+\frac{64}{(j+2)^{3}}-\frac{704}{3\,(j+2)^{2}} \nonumber \\
&-&\frac{3152}{9(j+2)}  
-32\,K(j)+64\,K(j+1)-64\,K(j+2) \,,  \nonumber \\
\gamma _{qq}^{(1)}(j)
&=&\frac{320}{9(j-1)}+\frac{32}{j^{3}}-\frac{96}{j^{2}}
+\frac{8}{3\,j}+\frac{88}{3(j+1)}  
-\frac{128}{3\,(j+2)^{2}}-\frac{608}{9(j+2)}-32\,K(j)  \nonumber \\
&+&32\,K(j+1)+{\hat{Q}}(j) \,,  \nonumber \\
\gamma _{q\varphi }^{(1)}(j)
&=&\frac{320}{9(j-1)}+\frac{32}{j^{3}}-\frac{96}{j^{2}}
+\frac{8}{3\,j}+\frac{32}{(j+1)^{2}}  
-\frac{64}{j+1}+\frac{64}{3\,(j+2)^{2}}+\frac{256}{9(j+2)}\nonumber \\
&-&32\,K(j)\,,
\nonumber \\
\gamma _{\varphi g}^{(1)}(j)
&=&\frac{64}{3(j-1)}+\frac{24}{j^{2}}-\frac{48}{j}
+\frac{72}{(j+1)^{2}}-\frac{236}{j+1}  
-\frac{48}{(j+2)^{3}}+\frac{176}{(j+2)^{2}}+\frac{788}{3(j+2)}  \nonumber\\
&-&48\,K(j+1)+48\,K(j+2) \,,  \nonumber \\
\gamma _{\varphi q}^{(1)}(j)
&=&\frac{64}{3(j-1)}+\frac{24}{j^{2}}-\frac{48}{j}
-\frac{22}{j+1}+\frac{32}{(j+2)^{2}}  
+\frac{152}{3(j+2)}-24\,K(j+1) \,,  \nonumber \\
\gamma _{\varphi \varphi }^{(1)}(j) &=&\frac{64}{3(j-1)}+\frac{24}{j^{2}}-
\frac{48}{j}-\frac{24}{(j+1)^{2}}+\frac{48}{j+1}  
-\frac{16}{(j+2)^{2}}-\frac{64}{3(j+2)}+{\hat{Q}}(j)  \nonumber
\end{eqnarray}
and in the polarized case (for odd $j$) 
\begin{eqnarray}
{\tilde{\gamma}}_{gg}^{(1)}(j)
&=&\frac{32}{j^{2}}-\frac{280}{3\,j}-\frac{32}{(j+1)^{3}}
+\frac{64}{(j+1)^{2}}+\frac{280}{3(j+1)} 
-32\,K(j)+32\,K(j+1)+{\hat{Q}}(j) \,,  \nonumber \\
{\tilde{\gamma}}_{gq}^{(1)}(j)
&=&\frac{16}{j^{2}}-\frac{140}{3\,j}-\frac{8}{(j+1)^{3}}
+\frac{32}{(j+1)^{2}}+\frac{142}{3(j+1)}  
-16\,K(j)+8\,K(j+1) \,,  \nonumber \\
{\tilde{\gamma}}_{qg}^{(1)}(j)
&=&\frac{-64}{j^{2}}+\frac{568}{3\,j}+\frac{64}{(j+1)^{3}}-\frac{128}{(j+1)^{2}}
-\frac{560}{3(j+1)}+32\,K(j)-64\,K(j+1) \,,  \nonumber \\
{\tilde{\gamma}}_{qq}^{(1)}(j)
&=&\frac{-32}{j^{2}}+\frac{284}{3\,j}+\frac{16}{(j+1)^{3}}
-\frac{64}{(j+1)^{2}} 
-\frac{284}{3(j+1)}+16\,K(j)-16\,K(j+1)+{\hat{Q}}(j) \,,  \nonumber
\end{eqnarray}
where
\begin{eqnarray}
&{\hat{Q}}(j)=-\frac{4}{3}\,{S_{1}(j)} 
+16\,{S_{1}(j)}\,{S_{2}(j)}+8\,{S_{3}(j)}-8\,{{\tilde{S}}_{3}(j)}
+16\,{\tilde{S}}_{1,2}(j)\, ,&  \\[3mm]
&\,K(j) =\displaystyle {\frac{1}{j}\left( \frac{S_{1}(j)}{j}+S_{2}(j)
+{\tilde{S}}_{2}(j)\right)}\, ,&\\
&S_{k}(j) =\displaystyle {\sum_{i=1}^{j}\frac{1}{i^{k}}}\,,&\\
&{\tilde{S}}_{k}(j) =\displaystyle {\sum_{i=1}^{j}\frac{(-1)^{i}}{i^{k}}}\,,& \\
&{\tilde{S}}_{k,l}(j) =\displaystyle {
\sum_{i=1}^{j}\frac{1}{i^{k}}{\tilde{S}}_{l}(i)}\,.&
\end{eqnarray}
The analytical continuation of functions $\gamma _{ab}^{(1)}(j)$ 
($a,b=g,q,\varphi $) and $\tilde{\gamma}_{ab}^{(1)}(j)$ ($a,b=g,q$) to the
complex values of $j$ can be done analogously to Refs.~\cite{KK,KL}. The
procedure of the analytic continuation together with a detailed description
of our method of calculations will be presented elsewhere.

The eigenvalues of the anomalous dimension matrices are given below
\begin{eqnarray}
&&\gamma _{I}^{(1)}(j)=\gamma _{+}^{(1)}(j)={\hat{Q}}(j-2) \,,
\label{gammaplnp} \\[2mm]
&&\gamma _{{II}}^{(1)}(j)=\gamma _{0}^{(1)}(j)={\hat{Q}}(j) \,,
\label{gamma0np} \\[2mm]
&&\gamma _{{III}}^{(1)}(j)=\gamma _{-}^{(1)}(j)={\hat{Q}}(j+2) \,,
\label{gammamnnp} \\[2mm]
&&\gamma _{{IV}}^{(1)}(j)={\tilde{\gamma}}_{+}^{(1)}(j)={\hat{Q}}(j-1) \,,
\label{gammaplp} \\[2mm]
&&\gamma _{{V}}^{(1)}(j)={\tilde{\gamma}}_{-}^{(1)}(j)={\hat{Q}}(j+1)\,.
\label{gammamnp}
\end{eqnarray}
In fact they coincide with the expressions predicted in Ref.~\cite{KL}.
Indeed, using the two-loop result
\begin{eqnarray}
&&\hspace{-5mm}\gamma _+(j)=\tilde{\gamma }_+(j-1)= \gamma _0(j-2)=
\tilde{\gamma }_-(j-3)=\gamma _-(j-4)\nonumber\\
&&\hspace{-4mm}=\gamma (j) =-\displaystyle{\frac{\alpha _{s}\,N_{c}}{\pi }}
\,S_{1}(j-2)+{\left( \!\frac{
\alpha _{s}N_{c}}{4\pi }\!\right) }^{2}{\hat{Q}}(j-2)
\end{eqnarray}
for the universal anomalous dimension $\gamma (j)$
we can redefine $\alpha
_{s}\rightarrow \alpha _{s}(1-\alpha_{s}N_c/(12\pi ))$ to remove in
${\hat{Q}}(j)$ the term proportional to $S_{1}(j)$ (note, however, that it is an
additional redefinition of $\alpha _{s}$ in comparison with the transition
from ${\overline{{\rm MS}}}$ to ${\overline{{\rm DR}}}$-scheme). After this
substitution the above universal function ${\hat{Q}}(j)$ in two loops
coincides with $16Q(j)$ from Ref.~\cite{KL}.

For the polarized case the
Dokshitzer relation is similar to original one (below 
$\gamma_{ab}^{(1)}(j)=\gamma_{ab}$ and ${\tilde\gamma}_{ab}^{(1)}(j)={\tilde
\gamma}_{ab}$)
\begin{equation}
{\tilde \gamma}_{gg}+\frac{1}{2}{\tilde \gamma}_{qg}={\tilde
\gamma}_{qq}+2{\tilde \gamma}_{gq} \label{polm}
\end{equation}
and we can find that
\begin{eqnarray}
{\tilde \gamma}_{gg}+\frac{1}{2}{\tilde \gamma}_{qg} &=&{\hat{Q}}(j-1) \,,
\label{qpm1} \\
{\tilde \gamma}_{gg}-2{\tilde \gamma}_{gq} &=&{\hat{Q}}(j+1) \,.
\label{qpp1}
\end{eqnarray}
There are three relations for the unpolarized case
\begin{eqnarray}
&\gamma _{gg}+\gamma _{qg}+\gamma _{sg}=\gamma _{gq}+\gamma _{qq}+\gamma
_{sq}=\gamma _{gs}+\gamma _{qs}+\gamma _{ss} \,,&  \label{unpolm1} \\[2mm]
&\gamma _{gg}-4\gamma _{gq}+3\gamma _{gs}=
-\displaystyle{\frac{\gamma_{qg}}{4}}+\gamma_{qq}
-\frac{3\gamma_{qs}}{4}
= \frac{\gamma _{sg}}{3}-\frac{4\gamma_{sq}}{3}+\gamma_{ss} \,, &
\label{unpolm2} \\[2mm]
&12\gamma _{gq}-12\gamma _{gs}+3\gamma _{qg}-3\gamma _{qs}+4\gamma
_{sg}-4\gamma _{sq}=0 &  \label{unpolm3}
\end{eqnarray}
and one can verify that
\begin{eqnarray}
\gamma _{gg}+\gamma _{qg}+\gamma _{sg} &=&{\hat{Q}}(j-2)\,,  \label{qnm2} \\[2mm]
\gamma _{gg}-4\gamma _{gq}+3\gamma _{gs} &=&{\hat{Q}}(j)\,,  \label{qn} \\[2mm]
\gamma _{gg}-\gamma _{gq}-\frac{\gamma _{sg}}{3}+\frac{\gamma _{sq}}{3}
&=&{\hat{Q}}(j+2)\,.  \label{qnp2}
\end{eqnarray}
A complete diagonalization of the above anomalous dimension matrices
corresponds to the use of a slightly modified basis of the multiplicatively
renormalized twist-2 operators in comparison with the leading order~\cite
{Dubna,KL} due to the breakdown of the superconformal invariance (cf.~\cite
{BM}). But Eqs.~(\ref{polm})-(\ref{qnp2}) are correct in all orders with the
replacement of ${\hat{Q}}(j-2)$ by the exact universal dimension $\gamma
(j)$.

Following the analysis in Ref.~\cite{KL} the $\alpha _{s}^{3}$ correction to the universal anomalous
dimensions $\gamma (j)$ will be constructed as soon as the corresponding QCD
anomalous dimensions will be calculated (see the recent papers~\cite{VMV}
and references therein).

Recently there was a great progress in the investigation of the N=4 SYM
theory in a framework of the AdS/CFT correspondence~\cite{2M} where the
strong-coupling limit $\alpha _{s}N_{c}\rightarrow \infty $ is described by
a classical supergravity in the anti-de Sitter space $AdS_{5}\times S^{5}$.
In particular, a very interesting prediction~\cite{15M} (see also~\cite{M})
was obtained for
the large-$j$ behavior of the anomalous dimension for twist-2 operators
\begin{equation}
\gamma (j)=a(z)\,\ln j \,,\qquad\qquad z=\frac{\alpha _{s}N_{c}}{\pi }
\end{equation}
in the strong coupling regime
(see Ref.~\cite{16M} for
asymptotic corrections):
\begin{equation}
\lim_{z\rightarrow \infty }a=-z^{1/2}+\frac{3\ln 2}{8 \pi}+{\cal O}
\left(z^{-1/2}\right) \,.  \label{1d}
\end{equation}
Here we took into account,
that in our normalization $\gamma (j)$ contains the extra factor
$-1/2$ in comparison with that in Ref.~\cite{15M}.

On the other hand, all anomalous dimensions $\gamma _{i}(j)$ and
$\tilde{\gamma}_{i}(j)$ ($i=+,0,-$) coincide at large $j$ and our results for
$\gamma (j)$
allow one
to find two first terms of the small-$z$ 
expansion of the coefficient $a(z)$
\begin{equation}
\lim_{z\rightarrow 0}\,\tilde{a}=-z+\frac{\pi ^2-1}{12}\,
z^2+...\,.
\end{equation}

To go from this expansion to the strong coupling regime we
perform a resummation of the perturbative result using a method
similar to the Pade approximation
and taking into account, that for
large $N_{c}$ the perturbation series has a finite
radius of
convergency. Namely, we present $\tilde{a}$ as a solution of the
simple algebraic equation
\begin{equation}
z=-\widetilde{a}+\frac{\pi ^2-1}{12}\,
\widetilde{a}^{2}\,.
\end{equation}
From this
equation the following large-$z$   
behaviour of $\tilde{a}$ is obtained:
\begin{equation}
\tilde{a}\approx -1.1632\,\,z^{1/2}+0.67647+{\cal O}
\left(z^{-1/2}\right)
\end{equation}
in a rather good agreement with 
Eq.~(\ref{1d})
based on
the AdS/CFT correspondence.  Note, that if we write for
$\tilde{a}$ the more general  equation
\begin{equation}
z^{n}=\sum_{r=n}^{2n}C_{r}\, \widetilde{a}^{r}\,,
\end{equation}
the
coefficients $C_{r}$ for $n\geq 3$ can be chosen in such way to
include all known information about $a$. For $n=2$ we can
impose on $\widetilde{a}$ apart from
a correct small-$z$ expansion also
the condition $a=-z^{1/2}$ for large-$z$ and as
a result the asymptotic correction ${\cal O}\left(1\right)$ will be
in a better agreement with~Eq.~(\ref{1d}).

Further, for $j\rightarrow 2$ due to the energy-momentum conservation 
\begin{equation}
\gamma (j)=(j-2)\,\gamma ^{\prime }(2)+...\,,
\end{equation}
where the coefficient $\gamma
^{\prime }(2)$ can be calculated from our results in two first orders of the
perturbation theory:
\begin{equation}
\gamma ^{\prime}(2)=-\frac{\pi ^2}{6}z+1.2158 \,z^2+...\,.
\end{equation}
Using the same method of resummation as we used above for
 $\widetilde{a}$, we obtain for large $z$
\begin{equation}
\gamma ^{\prime}(2)=-0.9071\,z^{1/2}+0.6768+...\,.
\end{equation}
Let us take into account, that in
this limit $\gamma =1/2+i\nu +(j-1)/2\rightarrow 1+(j-2)/2$ for the
principal series of unitary representations of the M\"{o}bius group
appearing in the BFKL equation~\cite{next}. Therefore we obtain for
large
$z$
\begin{equation}
j=2-1.1024 \,z^{-1/2}-0.2148\, z^{-1}\,.
\end{equation}
in an
agreement with the result, that the Pomeron in the strong coupling
regime coincides with the graviton ~\cite{polch,CTan}. The correction $\sim z^{-1/2}$
to the graviton spin $j=2$ coincides in form with that obtained in 
Ref.~\cite{polch} from the AdS/CFT correspondence and the coefficient in
front of $z^{-1/2}$ was not calculated yet. Note, that for the soft
Pomeron the correction is $\sim z^{-1}$ ~\cite{CTan}.

One can attempt to
calculate the intercept of the BFKL Pomeron also using  its
perturbative expansion in Ref.~\cite{KL01}
\begin{equation}
j-1=2.7726 z -5.0238 z^2\,.
\end{equation}
After the Pade resummation
we obtain in the strong coupling regime
$j\simeq   2.5301 - 0.8444 z^{-1} $ in an reasonable
agreement with the AdS/CFT
estimate (see~\cite{CTan}). Note, however, that in the upper orders of
the perturbation theory the BFKL equation should be modified by
including the contributions from multi-gluon components of the Pomeron
wave function.

In the conclusion we want to stress again, that the AdS/CFT
correspondence unified with a resummation procedure gives a
possibility to relate weak and strong coupling results.\\[10mm]
%
{\large \bf Acknowledgments.}\\[5mm]
We thank Yu. Makeenko for the
helpful discussions.
This work was supported by the Alexander von Humboldt Foundation fellowship
(A.V.K.), the RFBR grants 00-15-96610, 02-02-17513 and INTAS grant 00-366.

\end{document}